\documentclass[prl,aps,amssymb,nofootinbib,twocolumn,showpacs,floatfix]{revtex4}
\usepackage{amsmath}
\usepackage{amssymb}
\usepackage{amsthm}
\usepackage{amsfonts}
\usepackage{enumerate}
\usepackage{latexsym}
\usepackage[dvips]{graphicx}
\usepackage{float}
\usepackage{subfigure}

\newcommand{\beq}{\begin{equation}}
\newcommand{\eneq}{\end{equation}}
\newcommand{\beqs}{\begin{equation*}}
\newcommand{\eneqs}{\end{equation*}}

\input{epsf}

\begin{document}

\tolerance 1000

\title{Mott Fermionic ``Quantum'' Criticality Beyond
Ginzburg-Landau-Wilson}

\author { Zaira Nazario$^\star$ and David I. Santiago$^\dagger$ }

\affiliation{$\star$ Max Planck Institute for the Physics of Complex
systems, N\"othnitzer Str. 38, 01187 Dresden, Germany \\
$\dagger$ Instituut-Lorentz for Theoretical Physics, Universiteit Leiden,
P. O. Box 9506, NL-2300 RA Leiden, The Netherlands}

\begin{abstract}

\begin{center}

\parbox{14cm}{The Mott critical point between a metal and a correlated
insulator has usually been studied via density or spin density bosonic
mode fluctuations according to the standard Ginzburg-Landau-Wilson
phase transition paradigm.  A moment's reflection leads to increasing
doubts that such an approach should work as the transition is
nonmagnetic, voiding the relevance of spin density modes. Charge
density modes are irrelevant since the long range Coulomb
interaction leads to a large plasmon gap and their
incompressibility. In solidarity with these doubts, recent
measurements of the Mott critical point in low dimensional organic
materials yield critical exponents in violent disagreement with the
bosonic mode criticality lore. We propose that fermionic fluctuations
control the behavior of the Mott transition. The transition thus has
an intrinsic quantum aspect despite being a finite temperature phase
transition. We develop this hitherto unexplored physics, obtain
experimental predictions and find agreement with one of the novel
unexplained experimental exponents. We conclude that this Mott
transition corresponds to a new universality class of finite
temperature critical points that contains quantum effects and cannot
be accounted for by conventional Ginzburg-Landau-Wilson wisdom.}

\end{center}

\end{abstract}

\pacs{71.10.Fd, 71.10.Hf, 71.30.+h}

\date{\today}

\maketitle

Strong Mott correlation physics is at the center of some the most
spectacular and important phenomena in condensed matter physics:
quantum antiferromagnetism in low dimensions\cite{af}, high
temperature superconductivity (high $T_c$ cuprates)\cite{cup}, heavy
fermion behavior\cite{hf1,hf2}, interesting phase competition and
their concomitant quantum and classical phase transitions, etc. In
particular, a very remarkable experiment was published
recently\cite{kanoda}. The phase transition between a Mott insulating
phase and a metallic phase was measured in the quasi-2D organic
material $\kappa$-(BEDT-TTF)$_2$Cu[N(CN)$_2$]Cl (phase diagram in
Figure 1). The importance of the measured properties of
the Mott transition lie in its novel unexpected exotic nature.

The critical exponents measured in the experiment {\it are in violent
disagreement with those obtained from any of the universality classes
that follow from the Landau-Ginzburg-Wilson phase (LGW) transition
paradigm\cite{wilson, hertz}.} This conventional and usually successful
LGW understanding of phase transitions gets critical properties from
the fluctuations of order parameter bosonic modes.  Even though the
Mott transition occurs between a featureless correlated insulator and a
featureless metal, interesting approximate mappings to the liquid-gas
transition (Ising universality class) seem to lead to an LGW picture
of the transition\cite{crit1,crit2} driven by particle density modes.

In the Mott system, there are reasons to think that if an LGW picture
is to work, the particle density modes would be the main players. The
system consists of the fermionic quasiparticles and their quasi-bound
state spin and particle density modes.  Since there is no magnetism in
either of the phases, spin density modes play no rule. This seem to
single out density modes as the only possible candidates for LGW
physics for the Mott transition. On the other hand, complete
disagreement of predicted exponents from measured ones\cite{kanoda}
points to the irrelevance of density modes and toward the Mott
transition being a completely novel universality class.
   \begin{figure}[ht!]
      \label{phased}
      \includegraphics[width=3.5cm, angle=270]{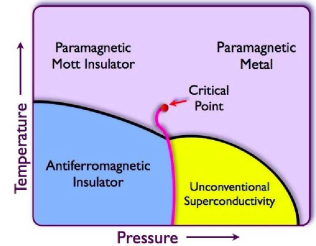}
      \caption{Pressure-temperature phase diagram of the family of
      compounds $\kappa$-(BEDT-TTF)$_2$X studied by F. Kagawa, {\it
      et. al.} on reference \cite{kanoda}. The measurements focused on
      the member $\kappa$-(BEDT-TTF)$_2$Cu[N(CN)$_2$]Cl.}
    \end{figure}

With the hindsight of the experimental disagreement with density mode
LGW physics, we can find important physical reasons as to why the
density modes should be irrelevant at the transition. In electronic
systems, proper determination of the effects of the long range Coulomb
interactions of mobile electrons in a background of very heavy
positive ions leads to a large plasmon gap, and thus to
incompressibility for long wavelength density modes, with a leftover
short range interaction between the electron-like and hole-like
quasiparticles in a metal\cite{plasma}. Such a short range interaction
justifies the neglect of the long range Coulomb interaction among
quasiparticles as it has been canceled via screening effects. Neglect
of such long range effects in the interactions between density modes
is incorrect as the plasmon gap is large and prevents the long
wavelength density modes from becoming soft and playing an important
role in the transition.

On the face of the experimental irrelevance of spin and particle
density bosonic modes as drivers of the transition, we identify the
quasiparticle fermionic degrees of freedom and their fluctuations as
the main actors driving and determining the Mott critical transition.
The fermionic quasiparticles are the only remaining modes in the
system\footnote{The density and spin density modes, while not driving
the transition, do play an indirect but not unimportant role as they
can mediate interactions between the fermionic quasiparticles. Without
such an interaction, no transition would be possible.}. that can play
a role at the transition.  We see that even though the Mott transition
we are considering is a finite temperature transition, fermionic
quasiparticles and thus their statistics, play an essential role at
the transition.

More importantly fermionic fields cannot acquire a nonzero macroscopic
average as no two of them can occupy the same quantum state; fermions
cannot be order parameters. Thus despite being a finite temperature
transition, the Mott transition contains an essential quantum aspect
to it stemming from the relevance of fermionic quasiparticles and
contains no order parameter physics. { \it Transitions of this type,
finite temperature but with an essential quantum aspect to them, are
completely novel, are not compatible with the Ginzburg-Landau-Wilson
paradigm\cite{hertz,wilson}.} They have hitherto never been studied.

In order to study more quantitatively the Mott transition, we start
from the partition function for the Hubbard model:
    \begin{align}
      \begin{aligned}
	Z &= \int \mathcal D \psi^\dagger_{\sigma}\mathcal D
	\psi_{\sigma} e^{-S} \\
	S &= \int_0^\beta d \tau d^2 x \left[\sum_{\sigma} \left(
	\psi^\dagger_{\sigma} \frac{\partial \psi_{\sigma}}{\partial
	\tau} - \frac{\mu}{\hbar} n_\sigma\right) \right. \\
	& \left. + \frac{\hbar}{2m} \sum_\sigma \nabla
	\psi_\sigma^\dagger \cdot \nabla \psi_\sigma + \frac{U
	a^2}{\hbar} n_{\uparrow} n_{\downarrow} \right] \\
	&= \int_0^\beta d \tau d^2 x \left[\sum_{\sigma} \left(
	\psi^\dagger_\sigma \frac{\partial \psi_\sigma}{\partial \tau}
	- \frac{\mu}{\hbar} \psi^\dagger_\sigma \psi_\sigma \right)
	\right. \\
	& \left. + \frac{\hbar}{2m} \sum_\sigma \nabla
	\psi_\sigma^\dagger \cdot \nabla \psi_\sigma + \frac{U
	a^2}{\hbar} \psi^\dagger_\uparrow \psi_\uparrow
	\psi^\dagger_\downarrow \psi_\downarrow \right] \; .
      \end{aligned}
    \end{align}
where $\sigma = \uparrow, \downarrow$ represents the spin and $a$ is
the lattice spacing. The Mott critical point will occur at critical
values $U = U_c$, $T = T_c$. The Hubbard interaction is somewhat
inconvenient to manage due to its quartic nature. It can be written
more conveniently with the help of a Hubbard-Stratonovich decoupling
leading to 
    \begin{align}
      \begin{aligned}
        \label{kanoda}        
	&Z = \int \mathcal D \lambda \mathcal D \rho \mathcal D \sigma
	\mathcal D \psi^\dagger_{\sigma}\mathcal D \psi_{\sigma}
	e^{-S} \\
	&S= \int_0^\beta d \tau d^2 x \left[\sum_{\sigma}
	\psi^\dagger_{\sigma}\frac{\partial \psi_{\sigma} }{\partial
	\tau} - \frac{\mu}{\hbar} n \right. \\
        &\left. \quad + \frac{\hbar}{2m} \sum_\sigma \nabla
        \psi_\sigma^\dagger \cdot \nabla \psi_\sigma + \frac{U a^2}{4
        \hbar} \sigma^2 + \frac{U a^2}{4 \hbar} \rho^2 \right. \\
        & \left.\qquad \qquad \qquad + \frac{U a^2}{2 \hbar} s \sigma
        + i \frac{U a^2}{2 \hbar} n \rho \right] \; ,
      \end{aligned}
    \end{align}
where we have defined $ n(\vec x) \equiv n_{\uparrow}(\vec x) +
n_{\downarrow}(\vec x)$ and $s(\vec x) \equiv n_{\uparrow}(\vec x) -
n_{\downarrow}(\vec x)$. The partition function is that of fermions
interacting with bosonic fields that correspond in a sense to particle
and spin density modes ($i \rho$ and $\sigma$). This is an exact
partition function for the Hubbard model and gives the correct
description of its different phases and phase transitions.

\begin{figure}[!ht]
      \includegraphics[width=2cm, angle=270]{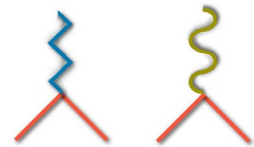}
      \caption{Feynman diagrams representing the interaction of
      fermions with spin density and density fluctuations. Straight
      lines represent fermions, zig-zag lines represent spin density
      fluctuations and wavy lines represent density fluctuations.}
      \label{vert}
    \end{figure}

\begin{figure}
      \includegraphics[width=0.85cm, angle=270]{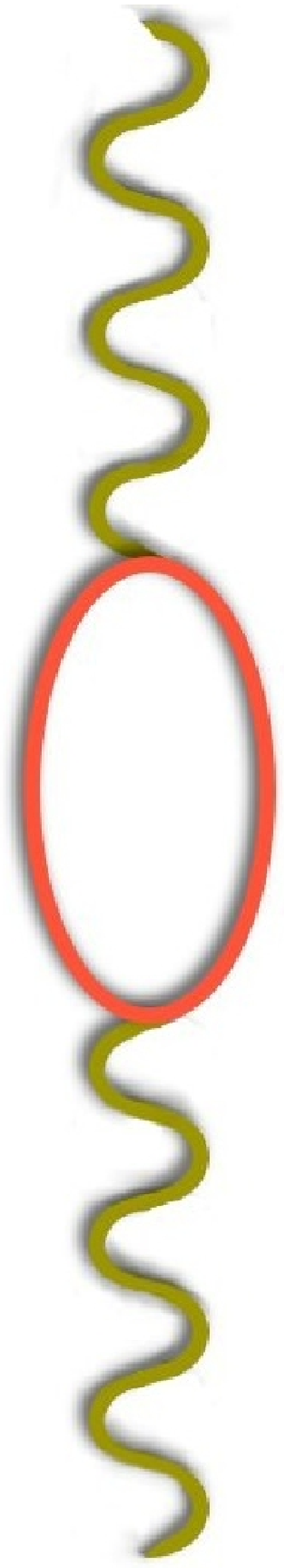}
      \includegraphics[width=0.85cm, angle=270]{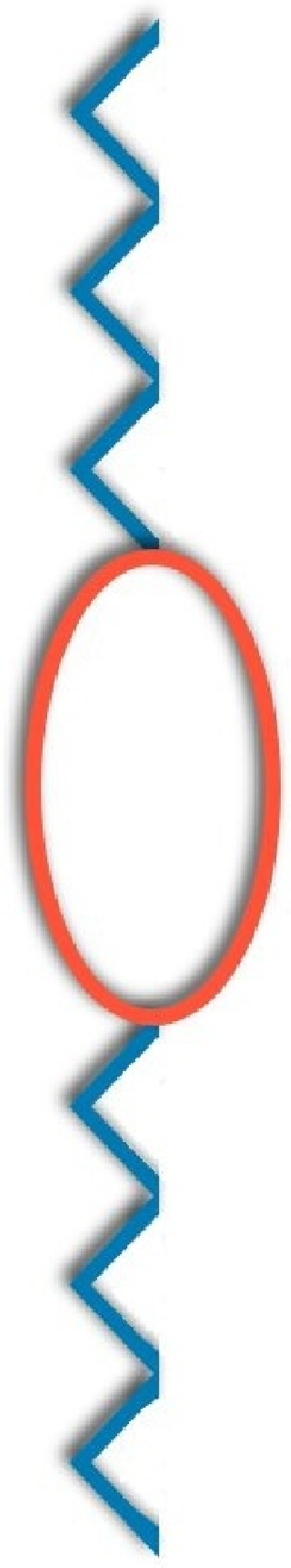}
      \caption{Fermion fluctuation responsible for generating density
      and spin density fluctuations dynamics. The solid line
      represents the fermions, the wavy lines are density fluctuations
      and the zig-zag lines are spin density fluctuations.}
      \label{rhoeff}
    \end{figure}

The Hubbard interaction generates an effective interaction between
particle and holes and spin density and density modes, as shown
diagrammatically in Figure \ref{vert}, given by

    \begin{align}
      \begin{aligned}
	&- \frac{T}{\hbar} \sum_n \int \frac{d^2 \vec k}{(2\pi)^2}
	\left[ i g_\rho n (\omega_n, \vec k) \rho (-\omega_n, -\vec k)
	\right. \\
	&+ \left. g_\sigma s(\omega_n, \vec k) \sigma (-\omega_n,
	-\vec k) \right]
      \end{aligned}
    \end{align}
where $g_\rho = g_\sigma = (U a^2)/(2 \hbar) = g_{rs}$ and bosonic
Matsubara frequency $\omega_n = 2 \pi n T/\hbar$. The dynamics of the
collective modes and their effective interactions with the
quasiparticles are equivalent to, and take the place of, the Hubbard
interaction.  The Hubbard interaction between fermions generates also
dynamics for density modes and spin density modes via fermionic
fluctuations. In fact, particle-hole fluctuations of the form
illustrated by the Feynman diagrams in Figure \ref{rhoeff} generate a
term in the Boltzmann weight of the partition function or effective
``Euclidean action'' $S$ describing the dynamics of the short range
density and spin density fluctuations induced by the onsite repulsion.

\begin{figure}[ht!]
      \includegraphics[width=2cm, angle=270]{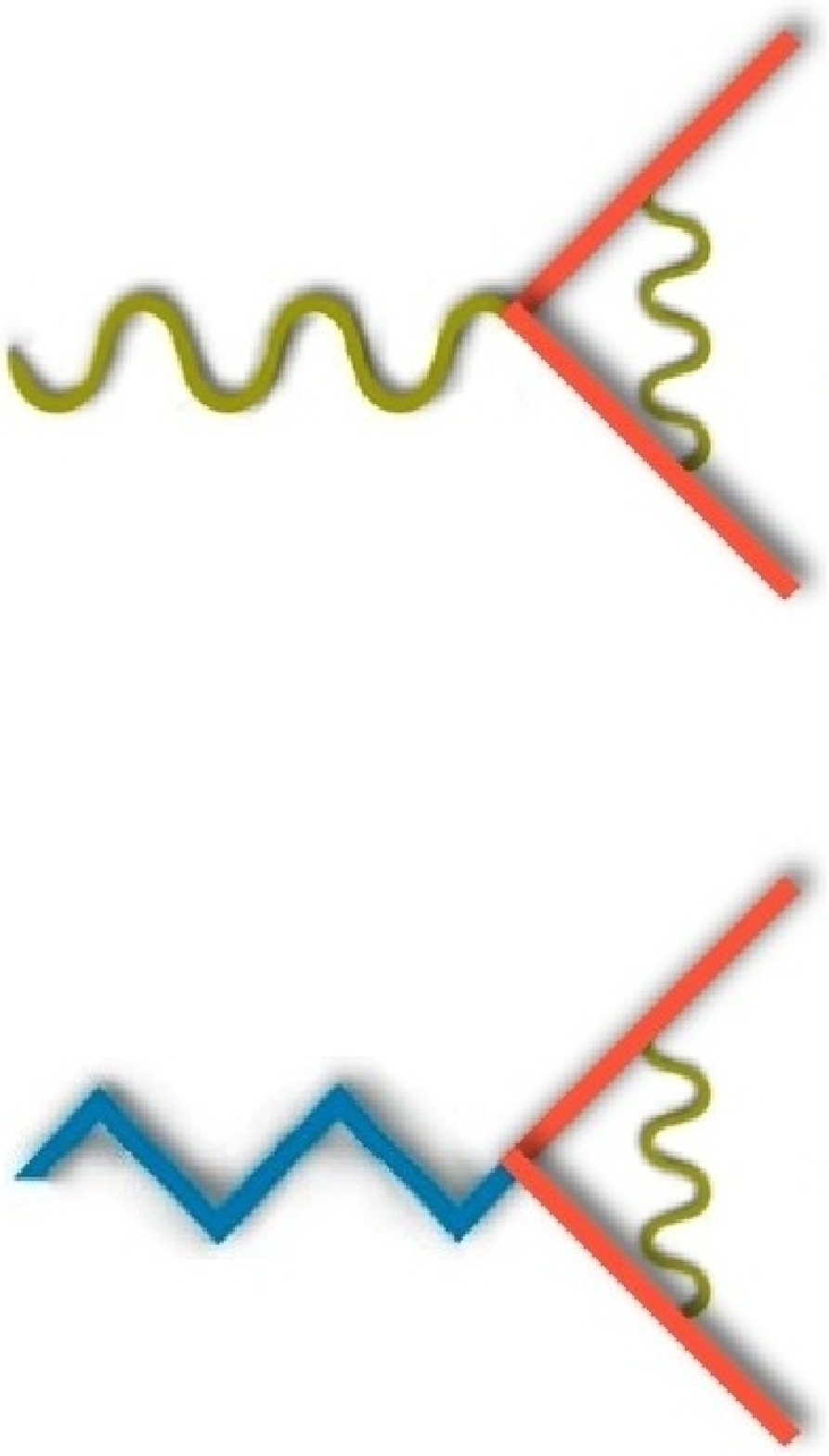}
      \includegraphics[width=2cm, angle=270]{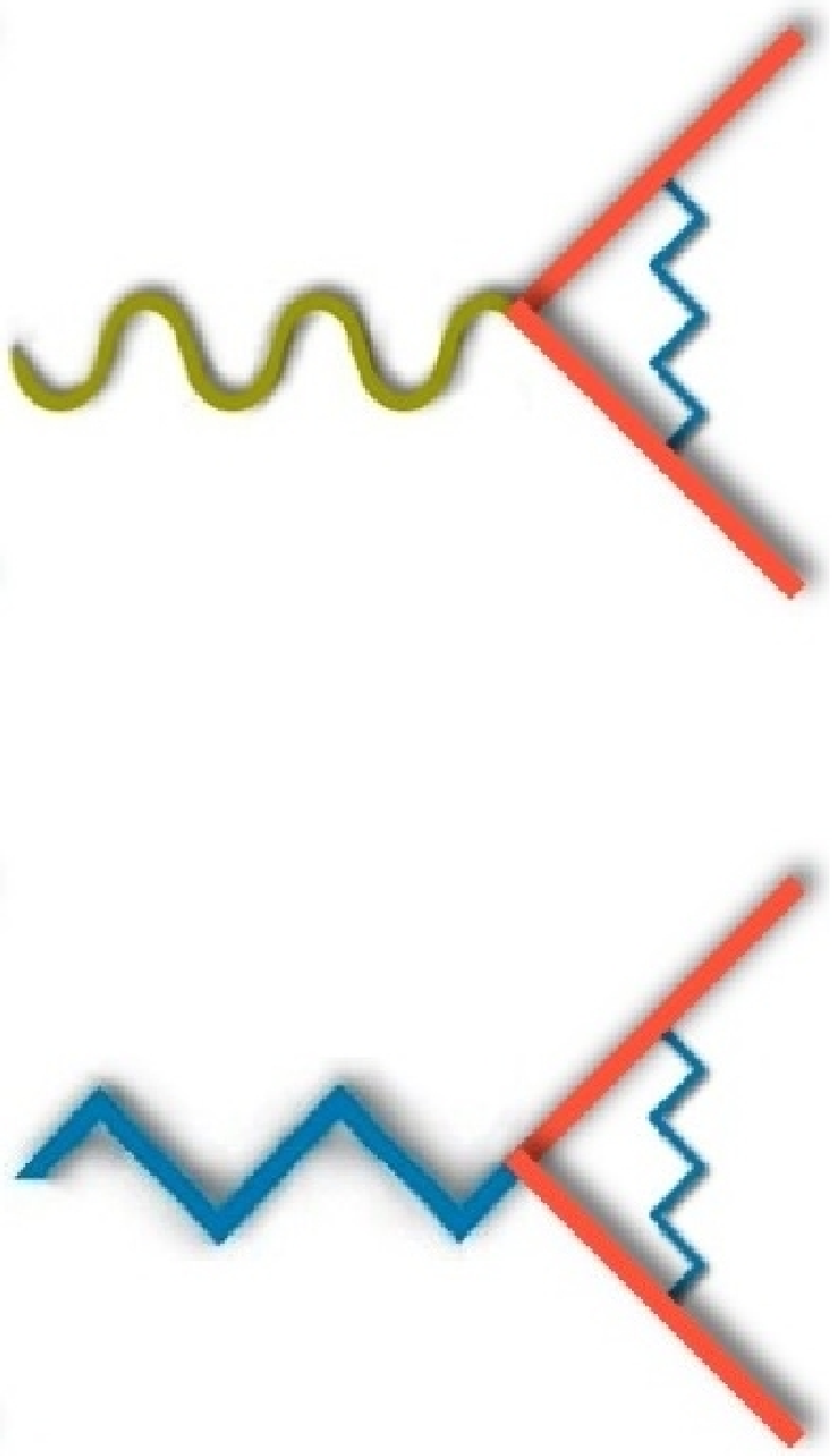}
      \caption{Feynman diagrams representing the renormalization of
      the interaction of fermions with spin density and density
      fluctuations. Straight lines represent fermions, zig-zag lines
      are spin density fluctuations and wavy lines are density
      fluctuations.}
      \label{vertren}
    \end{figure}

The effective theory of the system consists of metallic fermions,
density and spin density modes, and the interaction between these
collective modes and the fermions with strength $g_{rs}$. Single
particle or hole fluctuations will dress or renormalize the
interactions between the density modes and the fermions, either
enhancing or diminishing them. Processes of the form shown in Figure
\ref{vertren}
change the value of the effective coupling constant $g_{rs}$.  

When we
consider large momenta or small distance fluctuations, i.e. those
between the microscopic momentum cutoff $\Lambda$ (taken to be the inverse
lattice spacing $1/a$) and a lower cutoff $\mu$, processes of this
type give a renormalized coupling constant appropriate to the distance
scale $1/\mu$, 
    \beq
      g_{rs}^\mu = \frac{ g_{rs}^\Lambda}{1 - \frac{4}{3} \pi^4 \,
      \left( g_{rs}^\Lambda \right) ^2 \, \frac{\hbar^2 m v_F^2
      \Lambda^2}{T^3} \, \Lambda (\Lambda - \mu)} \;.
    \eneq
The thermodynamic long wavelength behavior follows from the small
momentum ($\mu \rightarrow 0$) behavior of the renormalized coupling
constant. For $\left( g_{rs}^\Lambda \right) ^2 < g_c^2 \equiv 3 T^3
/(4 \pi^4 \hbar^2 m v_F^2 \Lambda^4 S_4)$, the coupling constant is
finite and the system is in the weakly interacting metallic phase.
The divergence at $g_{rs} = g_c$ identifies one of the conditions for
the Mott transition: the divergence of the renormalized coupling
constant. The Mott critical point occurs for $U_c^2 = 4 \hbar^2
\Lambda^4 g_c^2= 3 T^3 / (\pi^4 m v_F^2)$.

\begin{figure}
      \includegraphics[width=1.2cm, angle=270]{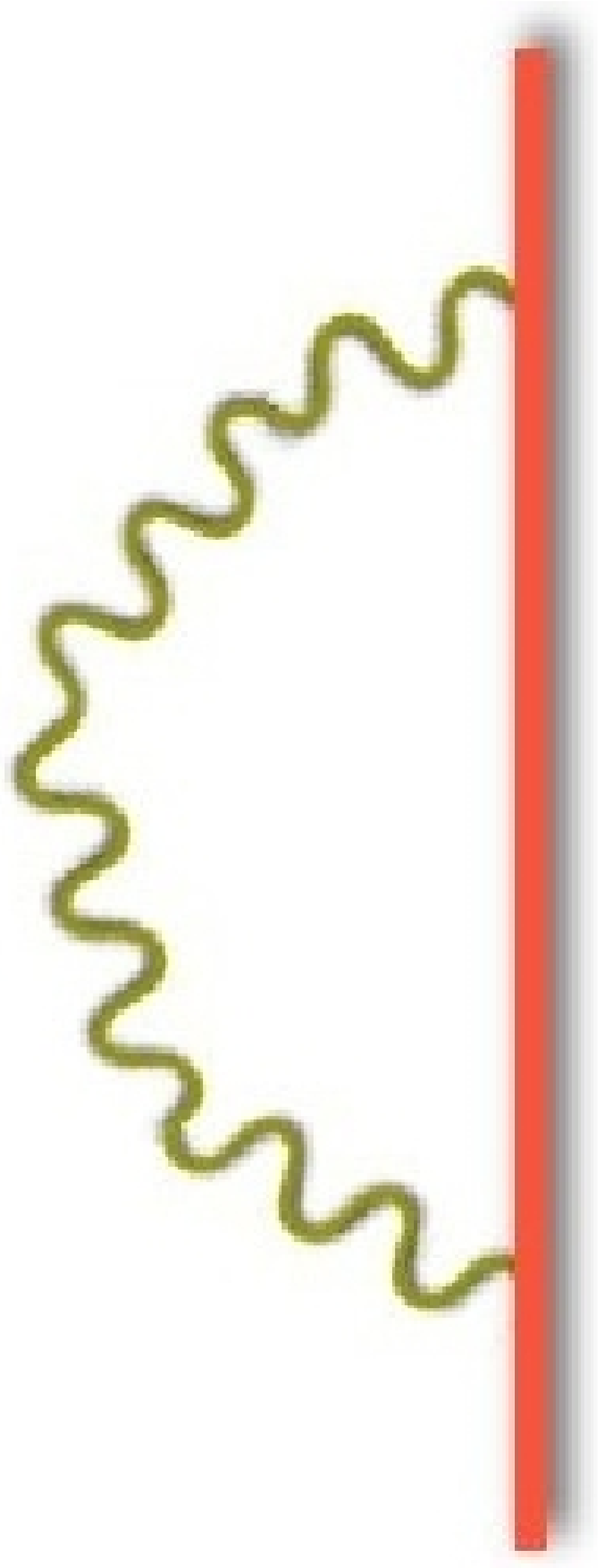}
      \includegraphics[width=1.2cm, angle=270]{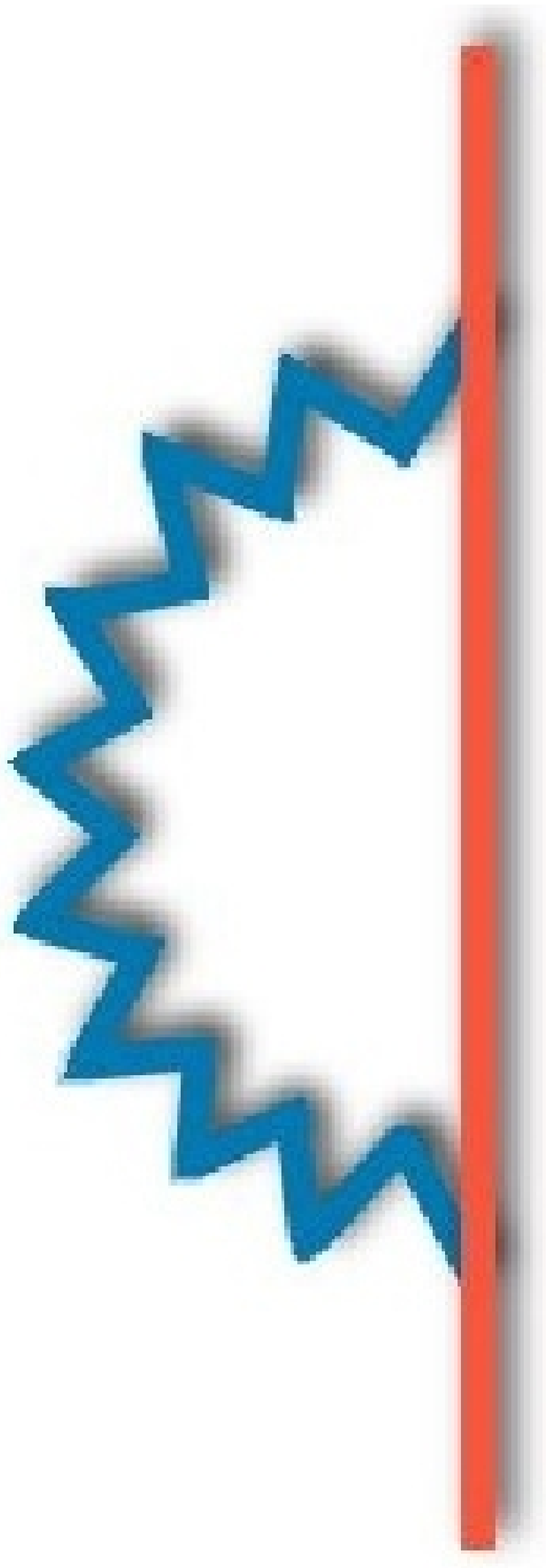}
      \caption{Feynman diagrams representing the self energy
      corrections to the quasiparticle propagator due to density and
      spin density fluctuations. Straight lines represent fermions,
      zig-zag lines are spin density fluctuations and wavy lines are
      density fluctuations.}
      \label{selfe}
    \end{figure}

The divergence of the renormalized coupling constant is not enough to
uniquely fix the Mott critical point. The missing ingredient is the
behavior of quasiparticle motion and dynamics. The quasiparticle
dynamics follows from its Green's function, which in the metallic
phase is $ G (\vec k, \omega_n) \, = (-i \omega_n + v_F k)^{-1}$, with
fermionic Matsubara frequency $\omega_n = (2 n+1) \pi T/ \hbar$. The
fermionic Green's function is renormalized by fluctuations of density
and spin density modes as illustrated in Figure \ref{selfe},
which provide a self energy correction to the fermion propagator
appropriate to the distance scale $1/\mu$
    \begin{align}
      \begin{aligned}
	\Sigma_\psi (\omega_n, \vec k) &\simeq i \omega_n \, g_{rs}^2
        \frac{2 \hbar^2 m v_F^2}{\pi^6 T^3} \, \Lambda^3 (\Lambda -
        \mu) \, \left[ \frac{\pi^4}{48} \right. \\ 
	&+ \left. \frac{U^2 m^2 v_F^4}{64 \pi^6 T^4} \, S_8 \right] \\
	S_8 &= \sum_n \frac{1}{(2 n + 1)^8} = \frac{17 \pi^8}{(315)
	2^8}
      \end{aligned}
    \end{align}
leading to a renormalized quasiparticle propagator $ G (\vec k,
\omega_n) \simeq (- i \omega_n + v_F k - \Sigma_\psi)^{-1}$.  This
renormalized propagator has standard form with a renormalized Fermi
velocity $v_F^R (\mu)$ and a renormalized quasiparticle weight $Z
(\mu)$
    \begin{align}
      \begin{aligned}
	\label{Gcrit}
	G (\vec k, \omega_n) &\simeq  \, Z
	(\mu) \frac{1}{\left[ - i \omega_n + v_F^R (\mu) \, k \right]} \\
	Z (\mu) &\simeq 1 - g_{rs}^2 \frac{m \hbar^2 v_F^2}{\pi^4
        T^3} \, \Lambda^3 (\Lambda - \mu) \, \left[ \frac{\pi^2}{24}
        \right. \\
	&+ \left. \frac{17 U^2 m^2 v_F^4}{(315) 2^{13} T^4} \right] \\
	v_F^R (\mu) &\simeq v_F \, \left\{ 1 - g_{rs}^2 \frac{m
        \hbar^2 v_F^2}{\pi^4 T^3} \, \Lambda^3 (\Lambda - \mu) \,
        \left[ \frac{\pi^2}{24} \right. \right. \\
	&+ \left. \left. \frac{17 U^2 m^2 v_F^4}{(315) 2^{13} T^4}
        \right] \right\} \;.
      \end{aligned}
    \end{align}
At long wavelengths, $\mu = 0$, we see that the renormalized Fermi
velocity can be zero for a critical value of $g_{rs}$
    \beq
      \label{gcritv}
      \left( g_{rs} \right)^2 = g_c^2 \equiv \frac{\pi^4 T^3}{m
      \hbar^2 v_F^2 \Lambda^4 \left[ \frac{\pi^2}{24} + \frac{17 U^2
      m^2 v_F^4}{(315) 2^{13} T^4} \right]} \;.
    \eneq
This corresponds to a Fermi surface instability as the electrons
immobilize and get localized. This vanishing of the Fermi velocity is
equivalent to a divergence of the renormalized mass as seem from
$v_F^R(\mu) = \hbar k_F/ m^R(\mu)$.  Furthermore the quasiparticle
weight $Z(\mu)$ vanishes at $\mu=0$ together with the vanishing of the
Fermi velocity.  The two conditions for the Mott critical point, the
vanishing of the renormalized Fermi velocity and divergence of the
renormalized coupling constant, need not be compatible.  Making them
consistent uniquely fixes the Mott critical point to occur at $T_c \propto U_c
\propto mv_F^2$.

If we fix the temperature and coupling constant to their critical
values, the renormalized Fermi velocity and spectral weight at momentum
scale $k$ are $ Z (k) \simeq k / \Lambda$ and $ v_F^R (k) \simeq v_F
\, k / \Lambda$ leading to a quasiparticle propagation amplitude $G
(\vec k, \omega_n) \simeq \, k / \left[ \Lambda ( - i \omega_n + v_F
k^2 / \Lambda ) \right]$ {\it Therefore we predict that at the novel
Mott critical point recently measured\cite{kanoda}, quasiparticle
dispersion as measured via photoemission experiments will correspond
to quadratically dispersing gapless fermions.} In fact, the retarded
Green's function to be extracted from photoemission is 
   \beq
G (\vec k,
\omega) = i \frac{1}{\omega + i \delta - v_F \frac{k^2}{\Lambda}}
\frac{1}{ e^{\hbar \beta v_F k^2 / \Lambda} + 1} \; .
   \eneq
  
The picture of the Mott transition that we have just obtained is that
of metallic fermionic quasiparticles linearly dispersing perpendicular
to the Fermi surface, which morph into gapless quadratically
dispersing fermions which open a gap continuously as the system is
tuned to the Mott insulating side. Furthermore $Z(0) \sim T-T_c$ and $Z(0) \sim
U-U_c$ near the Mott critical point. {\it We thus also predict that
the spectral weight at the Fermi level as measured by photoemission
will go to zero linearly in the temperature deviation from
criticality.}

One of the measured\cite{kanoda} novel exponents was that the
deviation of the conductance from the critical conductance scales
linearly with $T - T_c$. The conductivity of the system can be
obtained from density or current fluctuations, which follow from the
fermionic quasiparticles susceptibility via linear response theory
    \beq 
      \sigma(\omega, \vec k) \sim \frac{\omega_n}{k^2} m \left[ 1 -
      \frac{|\omega_n|}{\sqrt{\left(v_F k \right)^2 + \omega_n^2}}
      \right] \;.
    \eneq 
The non-dynamical conductivity or conductance is given by the long
wavelength and small frequency limits of the dynamical
conductivity. The scattering rate, or inverse mean free time,
introduces a frequency scale $\omega \sim 1/\tau$ below which the
physics described by the conductivity is vitiated and smaller
frequencies are unavailable. Similarly, the mean free path introduces
a momentum scale $k \sim 1/l$. The conductance is then
    \beq
      G = \sigma(1/\tau, 1/l) \sim \frac{l^2}{\tau} m \left[ 1 -
      \frac{1}{\sqrt{\left(v_F \tau / l\right)^2 + 1}} \right] \;.
    \eneq

In general the conductance is the sum of two conductances: that
originating from scattering of impurities and that originating from
scattering of phonons. Since impurities are fixed in space, such
scattering has a fixed mean free path $l$, with scattering time
subservient to $l$ via $\tau =l/v_F$. The contribution to the
conductance from impurities has the form $ G_{imp} \sim v_F m$. Since
the renormalizations of $v_F$ and $m$ cancel among themselves, this
conductance is both constant in the metallic phase and at the Mott
critical point. There is also a conductance arising from phononic
scattering. Scattering from phonons, being a dynamical process,
introduces an intrinsic scattering rate $1/\tau$ with the mean free
path enslaved to it via $l = v_F \tau$. Therefore, the conductance
arising from phonon scattering has the form $ G_{phn} \sim v_F^2 m
\tau$, which vanishes at the critical point as $T - T_c$.

Since the total conductance is $G = G_{imp} + G_{phn}$, the critical
conductance is constant, arising from impurity scattering and thus
nonuniversal. {\it We predict the critical conductance to be material
dependent}. The deviations from the critical conductance for
temperatures near $T_c$ arise from phonon scattering.  These
deviations from the critical conductance vanish linearly with $T- T_c$
in disagreement with conventional LGW wisdom, but in
agreement with experiment\cite{kanoda}.

We thus conclude that our development can account correctly for at
least one of the non-LGW exponents and thus points toward {\it the
Mott critical point corresponding to a new universality class that,
despite being finite temperature, is controlled by an essentially
quantum aspect, fermionic fluctuations, and falls out of the reach of
the Ginzburg-Landau-Wilson paradigm.}

\end{document}